\documentclass[11pt,a4paper]{article}

\usepackage[T1]{fontenc}
\usepackage[utf8]{inputenc}
\usepackage{lmodern}
\usepackage[margin=2.5cm]{geometry}
\usepackage{microtype}
\usepackage{amsmath,amssymb}
\usepackage{mathpartir}
\usepackage{booktabs,array}
\usepackage{float}
\usepackage{algorithm}
\usepackage[noend]{algpseudocode}
\usepackage{xcolor}
\usepackage{tikz}
\usetikzlibrary{positioning,fit,calc,arrows.meta}
\usepackage{pgfplots}
\pgfplotsset{compat=1.18}
\usepackage[round,authoryear]{natbib}
\usepackage[hidelinks]{hyperref}
\usepackage[capitalise,nameinlink]{cleveref}
\usepackage{enumitem}
\setlist{itemsep=1pt,topsep=3pt}

\newcommand{\figAbstractMachine}{%
\begin{tikzpicture}[
    unit/.style={draw, minimum width=30mm, minimum height=6mm, font=\small},
    grp/.style={draw, dashed, inner sep=3mm, rounded corners=1pt},
    lbl/.style={font=\small\bfseries, anchor=south west, inner sep=1pt},
    >=Stealth]
  \node[unit] (mc) at (0,0) {Coordinating core};
  \node[unit, below=2mm of mc] (mm) {Central memory};
  \node[grp, fit=(mc)(mm), label={[lbl]north west:Master}] (M) {};
  \node[unit] (w1c) at (-3.6,-3.2) {Computing core};
  \node[unit, below=2mm of w1c] (w1m) {Local memory};
  \node[grp, fit=(w1c)(w1m), label={[lbl]north west:Worker}] (W1) {};
  \node[unit] (w2c) at (3.6,-3.2) {Computing core};
  \node[unit, below=2mm of w2c] (w2m) {Local memory};
  \node[grp, fit=(w2c)(w2m), label={[lbl]north west:Worker}] (W2) {};
  \node at ($(W1.east)!0.5!(W2.west)$) {$\cdots$};
  \coordinate (b) at ($(M.south)+(0,-0.45)$);
  \draw (M.south) -- (b);
  \draw[->] (b) -| ([yshift=4mm]W1.north);
  \draw[->] (b) -| ([yshift=4mm]W2.north);
\end{tikzpicture}}

\newcommand{\figHierMachine}{%
\begin{tikzpicture}[
    nd/.style={draw, minimum height=6.5mm, minimum width=15mm, font=\small},
    >=Stealth, level distance=13mm]
  \node[nd] (r) at (0,0) {Master};
  \node[nd] (m) at (-2.2,-1.4) {Master};
  \node     (d1) at (0,-1.4) {$\cdots$};
  \node[nd] (w) at (2.2,-1.4) {Worker};
  \node[nd] (w1) at (-3.5,-2.8) {Worker};
  \node     (d2) at (-2.2,-2.8) {$\cdots$};
  \node[nd] (w2) at (-0.9,-2.8) {Worker};
  \foreach \t in {m,d1,w} \draw[->] (r.south) -- (\t.north);
  \foreach \t in {w1,d2,w2} \draw[->] (m.south) -- (\t.north);
\end{tikzpicture}}

\newcommand{\figClusterMachine}{%
\begin{tikzpicture}[
    nd/.style={draw, minimum height=6mm, font=\small, inner xsep=2mm},
    lv/.style={font=\small\itshape, anchor=west, text=black!70},
    >=Stealth]
  \node[lv] at (-6.2, 0)    {Machine};
  \node[lv] at (-6.2,-0.75) {InfiniBand};
  \node[lv] at (-6.2,-1.5)  {Node level};
  \node[lv] at (-6.2,-2.25) {Front-side bus};
  \node[lv] at (-6.2,-3.0)  {Core level};
  \node[nd] (r) at (0.6,0) {Root-master};
  \node[nd] (n1) at (-1.4,-1.5) {Node-master};
  \node[nd] (n2) at ( 2.6,-1.5) {Node-master};
  \node at (0.6,-1.5) {$\cdots$};
  \node[font=\scriptsize] at (0.6,-1.15) {16 nodes};
  \node[nd] (a1) at (-2.4,-3) {Worker};
  \node[nd] (a2) at (-0.4,-3) {Worker};
  \node at (-1.4,-3) {$\cdots$};
  \node[font=\scriptsize] at (-1.4,-2.65) {8};
  \node[nd] (b1) at (1.6,-3) {Worker};
  \node[nd] (b2) at (3.6,-3) {Worker};
  \node at (2.6,-3) {$\cdots$};
  \node[font=\scriptsize] at (2.6,-2.65) {8};
  \foreach \t in {n1,n2} \draw[->] (r) -- (\t);
  \foreach \t in {a1,a2} \draw[->] (n1) -- (\t);
  \foreach \t in {b1,b2} \draw[->] (n2) -- (\t);
\end{tikzpicture}}

\definecolor{cP1}{RGB}{60,60,60}
\definecolor{cP2c}{RGB}{0,114,178}
\definecolor{cP2n}{RGB}{86,180,233}
\definecolor{cP8a}{RGB}{213,94,0}
\definecolor{cP8b}{RGB}{230,159,0}
\definecolor{cP16a}{RGB}{0,158,115}
\definecolor{cP16b}{RGB}{110,190,100}
\definecolor{cDown}{RGB}{0,114,178}
\definecolor{cUp}{RGB}{213,94,0}

\pgfkeys{/pgf/number format/1000 sep={\,}}
\pgfplotsset{
  sglbase/.style={
    width=0.92\linewidth, height=0.56\linewidth,
    grid=major, grid style={black!12},
    axis lines*=left, tick align=outside,
    tick label style={font=\footnotesize}, label style={font=\small},
    legend style={font=\footnotesize, draw=none, fill=none},
    every axis plot/.append style={line join=round},
  },
  sglperf/.style={sglbase,
    width=0.82\linewidth, height=0.44\linewidth,
    xlabel={Data size ($\times 10^3$ floating-point numbers)},
    ylabel={Time ($\mu$s)},
    x filter/.expression={x/1000},
    xmin=0, ymin=0,
    scaled x ticks=false, scaled y ticks=false,
    legend columns=3,
    legend style={at={(0.5,-0.3)}, anchor=north, /tikz/every even column/.append style={column sep=6pt}},
    legend cell align=left,
  },
}

\newcommand{\perfplot}[3]{%
\begin{tikzpicture}
\begin{axis}[sglperf, xmax=#2, ymax=#3, restrict y to domain=0:#3]
  \addplot[cP1,  thick] table {data/#1_p1.dat};    \addlegendentry{1 proc}
  \addplot[cP2c, thick] table {data/#1_p2c.dat};   \addlegendentry{2 procs (2 cores)}
  \addplot[cP2n, thick] table {data/#1_p2n.dat};   \addlegendentry{2 procs (2 nodes)}
  \addplot[cP8a, very thick, dashed] table {data/#1_th8.dat};  \addlegendentry{Predicted, 8 procs}
  \addplot[cP8a, thick] table {data/#1_p8a.dat};   \addlegendentry{8 procs ($2\times4$)}
  \addplot[cP8b, thick] table {data/#1_p8b.dat};   \addlegendentry{8 procs ($4\times2$)}
  \addplot[cP16a, very thick, dashed] table {data/#1_th16.dat}; \addlegendentry{Predicted, 16 procs}
  \addplot[cP16a, thick] table {data/#1_p16a.dat}; \addlegendentry{16 procs ($4\times4$)}
  \addplot[cP16b, thick] table {data/#1_p16b.dat}; \addlegendentry{16 procs ($8\times2$)}
\end{axis}
\end{tikzpicture}}

\newcommand{\figGapMPI}{%
\begin{tikzpicture}
\begin{axis}[sglbase, height=0.5\linewidth,
    xlabel={Number of processes}, ylabel={$g$ (ns per 32-bit word)},
    xmin=0, xmax=136, ymin=1.2, ymax=3.2,
    xtick={0,16,32,48,64,80,96,112,128},
    legend pos=north west, legend cell align=left]
  \addplot[cDown, thick, mark=*, mark size=1.6pt] coordinates
    {(2,1.38)(4,1.69)(8,1.89)(16,2.04)(32,2.14)(64,2.63)(96,2.88)(128,3.01)};
  \addlegendentry{$g_{\downarrow}$ (\texttt{MPI\_Scatterv})}
  \addplot[cUp, thick, mark=square*, mark size=1.5pt] coordinates
    {(2,2.15)(4,2.00)(8,2.05)(16,2.09)(32,2.09)(64,2.11)(96,2.13)(128,2.77)};
  \addlegendentry{$g_{\uparrow}$ (\texttt{MPI\_Gatherv})}
\end{axis}
\end{tikzpicture}}

\newcommand{\speedupplot}[3]{%
\begin{tikzpicture}
\begin{axis}[sglbase, width=0.9\linewidth, height=0.78\linewidth,
    xmin=0, xmax=#2, ymin=0, ymax=#2,
    xlabel={#3}, ylabel={Speed-up},
    legend pos=north west, legend cell align=left]
  \addplot[black!55, thick, domain=0:#2, samples=2] {x};
  \addlegendentry{Ideal}
  \addplot[cP8a, only marks, mark=triangle*, mark size=2.6pt] coordinates {#1};
  \addlegendentry{Measured}
\end{axis}
\end{tikzpicture}}

\newcommand{\vv}[1]{\overrightarrow{#1}}        
\newcommand{\ww}[1]{\widetilde{#1}}             
\newcommand{\kw}[1]{\mathbf{#1}}
\newcommand{\cfg}[2]{\langle #1,\ \sigma#2\rangle}
\newcommand{\sort}[1]{\mathbf{#1}}
\newcommand{\gdown}{g_{\downarrow}}
\newcommand{\gup}{g_{\uparrow}}
\newcommand{\kdown}{k_{\downarrow}}
\newcommand{\kup}{k_{\uparrow}}
\newcommand{\transfer}{\mathrel{:\!\sim}}
\newcommand{\numChd}{\mathit{numChd}}
\newcommand{\Cost}[1]{\hfill\makebox[4.2cm][l]{$\Leftarrow\ #1$}}

\algblockdefx{ParDo}{EndParDo}{\textbf{par do}}{\textbf{end par}}
\algnewcommand{\Gather}[2]{\textbf{gather} #1 \textbf{to} #2}
\algnewcommand{\Scatter}[2]{\textbf{scatter} #1 \textbf{to} #2}
\algnewcommand{\IfMaster}{\If{\textbf{master}}}
\algnewcommand{\Proc}[2]{\Statex #1(#2)\Statex\textbf{begin}}
\algnewcommand{\ProcEnd}{\Statex\textbf{end}}

\title{\textbf{SGL: Towards a Bridging Model for Heterogeneous Hierarchical Platforms}}
\author{%
  Chong Li$^{1,2,}$\thanks{Corresponding author: \href{mailto:chong.li@ieee.org}{\texttt{chong.li@ieee.org}}; ORCID: \href{https://orcid.org/0000-0002-4160-7170}{\texttt{0000-0002-4160-7170}}}
  \qquad Ga\'etan Hains$^{1,2}$\\[6pt]
  \small $^1$Laboratoire d'Algorithmique, Complexit\'e et Logique (LACL)\\
  \small Universit\'e Paris-Est Cr\'eteil, 61 avenue du G\'en\'eral de Gaulle, 94010 Cr\'eteil, France\\
  \small $^2$EXQIM S.A.S., 24 rue de Caumartin, 75009 Paris, France\\[2pt]
  \small (affiliations at the time of the work)
}
\date{}

\begin{document}
\maketitle

\begin{abstract}
\noindent
This paper introduces the Scatter-Gather parallel-programming and parallel-execution model in the form of a simple imperative Scatter-Gather Language (SGL). Its design is based on past experience with bulk-synchronous parallel (BSP) programming and BSP language design. SGL's novel features are motivated by the last decade's move towards multi-level and heterogeneous parallel architectures involving multi-core processors, graphics accelerators and hierarchical routing networks in the largest multiprocessing systems. The design of SGL is coherent with L.~Valiant's Multi-BSP while offering a programming interface that is even simpler than the primitives of Bulk-Synchronous Parallel ML (BSML). SGL appears to cover a large subset of all BSP algorithms while avoiding complex message-passing programming. Like all BSP-inspired systems, it supports predictable, portable and scalable performance. Moreover, SGL's explicit data distribution will allow automatic or programmable load-balancing.

\end{abstract}

\begin{center}
\fbox{\begin{minipage}{0.92\linewidth}\small
\textbf{Publication note.} This is an author-prepared version, based on the accepted manuscript, of the article
C.~Li and G.~Hains, ``SGL: towards a bridging model for heterogeneous hierarchical platforms'',
\emph{International Journal of High Performance Computing and Networking}, 7(2):139--151, 2012,
\copyright~Inderscience Enterprises Ltd.
The definitive version is available at \url{https://doi.org/10.1504/IJHPCN.2012.046371}.
\end{minipage}}
\end{center}

\section{Introduction}

Parallel programming and data-parallel algorithms have been the main techniques supporting high-performance computing for many decades. Like all non-functional properties of software, the conversion of computing resources into scalable and predictable performance involves a delicate balance of abstraction and automation with semantic precision.

With these goals in mind we apply the notion of explicit processes, because parallel speed-up cannot be part of programming semantics without an explicit notion of the number of physical parallel processors. A language for parallel algorithms must not express a function from computation events to physical units (which may not be injective) but just the inverse.

For bridging the gap between parallel systems and parallel algorithms, a major conceptual step was taken by \citet{Valiant1990}, who introduced his bulk-synchronous parallel (BSP) model. Inspired by complexity theory's PRAM model of parallel computers, Valiant proposed that parallel algorithms be designed and measured by accounting not only for the classical balance between time and parallel space (the number of processors) but also for communication and synchronisation. A BSP computation is a sequence of so-called \emph{supersteps}. Each superstep combines asynchronous local computation with point-to-point communications that are coordinated by a global synchronisation to ensure coherence and determinism. The resulting performance model is both realistic and tractable, so that researchers like \citet{McColl1998} were able to define BSP versions of all important PRAM algorithms, implement them and verify their portable and scalable performance as predicted by the model. BSP is thus a bridging model relating parallel algorithms to hardware architectures. A well-known implementation of BSP is the Paderborn University BSP library, or PUB \citep{Bonorden2003}. This work has been extended by many similar models: LogP \citep{Culler1993}, LogGP \citep{Alexandrov1995} for long messages, HLogGP \citep{Bosque2004} for heterogeneous processors, LogGPO \citep{Chen2009} to account for computation-communication overlap, and the CGM model \citep{Dehne1993}.

\citet{Loulergue2000calculus} designed BS-$\lambda$ as a minimal model of computation with BSP operations (an extension of the $\lambda$-calculus). BS-$\lambda$ became the basis for Bulk-Synchronous Parallel ML, BSML \citep{Loulergue2000thesis,Loulergue2010bsml}, a variant of Caml. BSML offers a purely functional semantics and a simple programming interface of only four operations: \texttt{mkpar} to construct parallel vectors indexed by processors, \texttt{proj} to map them back to lists/arrays, \texttt{apply} to generate asynchronous parallel computation, and \texttt{put} to generate communication and global synchronisation from a two-dimensional processor-processor pairing. As a result, parallel performance mathematically follows from program semantics and the BSP parameters of the host architecture.

By comparison with BSML, our new Scatter-Gather Language (SGL) model can be understood as:
\begin{enumerate}
  \item a generalisation to hierarchical architectures, by breaking the restriction to a flat vector of processors;
  \item a simplified set of primitives, where the ``many-to-many'' processor relations of \texttt{put} are replaced by simpler ``one-to-all'' and ``all-to-one'' relations.
\end{enumerate}
Our presentation of SGL is in the form of an imperative language, while BSML is a variant of ML, hence a functional language. But that last difference is less important than the above two.

Among the earlier works on BSML, the grid or departmental meta-computing ML (DMML) model and language of \citet{Gava2005} described a two-level BSP. Its lower level represents local parallel clusters and its top level a grid of such systems. SGL is both a generalisation of this design, for systems that have potentially more levels, and a restriction of the programming interface, by avoiding the most general communication operator \texttt{put}. SGL can cover lower levels inside the machines and also higher levels, for example from multi-core processors like the Nvidia Tesla or the STI Cell to giant systems like Tera-100 or Blue Gene. An analysis of the exact differences between both models remains to be done, but the experience gained on DMML will be reused for investigating and developing SGL.

SGL is well adapted to BSP-like algorithms whose communication patterns are all-to-all maps, but we believe that all BSP algorithms can be programmed in SGL for the following reasons. SGL can trivially emulate any BSP communication phase by simply concentrating all messages (gather), then reordering them sequentially before sending them to their destinations (scatter). But a random ``horizontal'' communication pattern could also be implemented natively after having been written as a gather-reorder-scatter sequence and then automatically converted into \texttt{put} or send-receive operations. This kind of optimisation is beyond the scope of this paper, but we have begun analysing its possibility in the language semantics.

SGL models parallel machines more precisely than BSP. SGL parameters take into consideration both the machine's computation performance and its communication performance. SGL's cost model can inform developers whether or not the root of a hierarchical tree-structured machine, such as a GPU cluster, can be a bottleneck. SGL can also help hardware vendors measure whether a new technology can be useful for a particular machine. For example, the well-known hierarchical tree-structured parallel machine CM-5 \citep{Leiserson1992}, using a fat-tree \citep{Leiserson1985} architecture for its data network, had good performance according to SGL's communication parameters.

The next section gives precise arguments that motivate our SGL model, in the context of existing similar models like BSP or LogP. \Cref{sec:model} then describes the SGL model's abstract machine, execution model and cost model, relating machine parameters to actual algorithm performance. \Cref{sec:prog} formally defines the programming model as the operational semantics of an imperative mini-language. \Cref{sec:exp} presents and analyses performance-measurement experiments for primitive algorithms. Finally, \cref{sec:concl,sec:future} conclude with a summary of ongoing work and open problems related to SGL.

\section{Motivations}\label{sec:motiv}

While BSML was evolving and practical experience with BSP algorithms was accumulating, one of its basic assumptions about parallel hardware was changing. The flat view of a parallel machine as a set of communicating sequential machines remains true but is more and more incomplete. Recent supercomputers like Blue Gene/L \citep{Almasi2003} and Blue Gene/P \citep{IBM2008} feature multi-processors on one card, multi-core processors on one chip, multiple-rack clusters, etc. The Cell/B.E. \citep{Kahle2005,Johns2007}, the Cell-based Roadrunner \citep{Barker2008} and GPUs feature a master-worker architecture. Moreover, \citet{Kumar2005} observe that heterogeneous chip multiprocessors present unique opportunities for improving system throughput and reducing processor power. The trend towards green computing puts even more pressure on the optimal use of architectures that are not only highly scalable but hierarchical and non-homogeneous.

During the 2000s it became obvious that models like BSP should adapt or generalise to the new variety of architectures. Yet programming simplicity and performance portability should be retained as much as possible. With these goals in mind, Valiant introduced Multi-BSP \citep{Valiant2011}, a multi-level variant of the BSP model, and showed how to design scalable and predictable algorithms for it; \citet{Stewart2011} proposed a programming model for BSP with partitioned synchronisation. The main new feature of Multi-BSP is its hierarchical nature, with nested levels that correspond to the natural layers of physical architectures. In that sense it preserves the notion of explicit processes and, as we will show in this paper with our SGL design, allows us to solve three pending problems with BSP programming:
\begin{enumerate}
  \item The flat nature of BSP is not easily reconciled with divide-and-conquer parallelism \citep{Hains1998}, yet many parallel algorithms (e.g., Strassen matrix multiplication, quad-tree methods) are highly artificial to program any other way than recursively.
  \item BSP was designed natively to scale out with the number of processors, and it was assumed that CPU clock speeds would continue to improve. However, there are barriers to further significant improvements in operating frequency due to voltage leakage across internal chip components and heat dissipation limits \citep{Sodan2010}. As a result, parallelism is nowadays used to scale up computing power, through multithreaded cores, multi-core CPUs, Cell processors, GP-GPUs, etc.
  \item We do not know how to design algorithms for nested-level systems with the original flat BSP model. Hierarchical architectures share communication resources inside every level but not between different levels. The BSP cost model is not suitable for this kind of architecture.
\end{enumerate}

Based on all the above developments and the last three observations, we now define the SGL model, which realises a programming model for Multi-BSP, generalises it to heterogeneous systems and yet simplifies the BSML primitives to a set of three. SGL assumes a tree-structured machine and uses only the following parallel primitives: \emph{scatter} (to send data from a master to its sub-machines), \emph{pardo} (to request asynchronous computations from the sub-machines) and \emph{gather} (to collect data back to the master). We show how to embed this model in an imperative language, how to define its operational semantics and its performance model, and how to use it to code useful algorithms (reduction, parallel scan and sorting). We then provide some initial measurements to validate its performance model, and explanations as to how it can easily adapt to heterogeneous systems.

The work summarised here is initial evidence that SGL can provide both a simpler programming interface than BSML and a richer execution model, like Multi-BSP. Like BSML, SGL has explicit data layout: its semantics uniquely determines data distribution on the nodes of the architecture tree. This allows for automatic tuning on most types of modern architectures in the following sense. Exact communication-computation overlap is not expressible in a language like SGL, but granularity (resp.\ load balancing) is under program control because communication (resp.\ data layout) is explicit. This feature is neither tested nor described in this paper, but the existing literature on BSP programming makes us confident that it will be retained by SGL.

\section{The SGL model}\label{sec:model}

The PRAM model \citep{Fortune1978} of parallel computation begins with a set of $p$ sequential von Neumann machines (processors). It then connects them with a shared memory and an implicit global controller (as in a SIMD architecture). The BSP model \citep{Valiant1990} relaxes the global-control and direct-memory-access hypotheses. It also begins with a set of $p$ sequential machines but assumes asynchronous execution, except for global synchronisation barriers and point-to-point communications that complete between two successive barriers. It is common to assume that the PRAM/BSP processors are identical, but it is relatively easy to adapt algorithms to a set of heterogeneous processors with varied processing speeds. In both cases the set of processors has no structure other than its numbering from $0$ to $p-1$.

This absence of structure on the set of processors can be traced back to the failure of a trend of research popular in the 1980s: algorithms for specific interconnect topologies, such as hypercube algorithms, systolic algorithms, rectangular-grid algorithms, etc. We trace this failure to the excessive variety of algorithms and architectures without a model to bridge the portability gap between algorithms and parallel machines. The SGL model we propose does introduce a degree of topology on the set of processors. But this topology is purely hierarchical and not intended to become an explicit factor of algorithm diversity. Just as PRAM/BSP algorithms use the $p$ parameter to adapt to systems of all sizes, our SGL algorithms use the machine parameters to adapt to all possible systems.

\subsection{The SGL computer or abstract machine}

\begin{figure}[htbp]
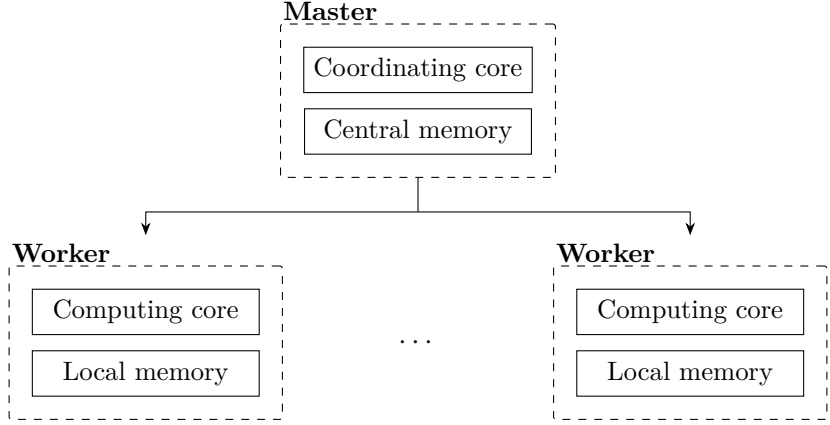

  \centering\figAbstractMachine
  \caption{SGL abstract machine.}\label{fig:abstract}
\end{figure}

The SGL model (\cref{fig:abstract}) defines a set of sequential processors, each composed of a computation element (``core'') and a memory unit.

\begin{figure}[htbp]
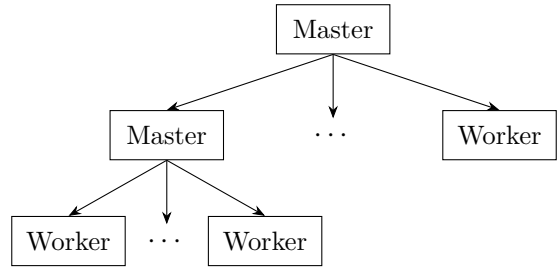

  \centering\figHierMachine
  \caption{SGL hierarchical machine.}\label{fig:hier}
\end{figure}

The processors are arranged in a tree structure, whose root is called a \emph{master} and whose children are either masters themselves or \emph{leaf-workers}. The number of children is not limited, so that the BSP/PRAM concept of a flat $p$-vector of processors is easily simulated in SGL. Different forms are possible:
\begin{itemize}
  \item only one worker without master: a sequential machine;
  \item a master with workers: a parallel machine, e.g., a BSP computer;
  \item a hierarchical machine (\cref{fig:hier}) of any shape.
\end{itemize}

The logical structure of an SGL computer satisfies the following constraints:
\begin{itemize}
  \item a system has one and only one root-master;
  \item a master coordinates its children through communication;
  \item a worker is controlled by one and only one master (masters can be replicated by underlying libraries for fault tolerance);
  \item communication is always between a master and its children.
\end{itemize}

\subsection{Execution model}

An SGL program execution is a sequence of supersteps. The initial data and the final result can be either distributed among the workers or centralised in the root-master. Each superstep is composed of four phases:
\begin{enumerate}
  \item a \emph{scatter} communication phase initiated by the master;
  \item an asynchronous computation phase performed by the children (a child's computation phase can itself be a sequence of supersteps);
  \item a \emph{gather} communication phase centred on the master;
  \item a local computation phase on the master.
\end{enumerate}
With these four phases, one can usually write an SGL program recursively:

\begin{algorithmic}
  \If{node is a master and its children are to be used}
    \State split input data into blocks;
    \State scatter data blocks;
    \State compute on individual data blocks in children (in parallel);
    \State gather children's results;
    \State combine children's results in master (if necessary);
  \Else
    \State compute on local data on node;
  \EndIf
\end{algorithmic}

\noindent The choice of using (or not) children in the recursive program depends on performance parameters that combine the (known) communication costs, synchronisation costs, computation costs and load balancing.

\subsection{Cost model}

Like all such models, SGL's main goal of expressing parallel algorithms requires a precise notion of the execution time of a program. We give here the mathematical form of this cost model, with the understanding that a full definition should be based on the operational semantics and will be given in a further document. The cost equations below are nevertheless sufficient for informal algorithm design and understanding, and for comparison with other parallel models.

The cost of an SGL algorithm (i.e., of its execution on given input data) is the sum of the costs of its supersteps. That follows from the understanding that supersteps are executed sequentially. The cost of an individual superstep is split into two independent terms, computation cost and communication cost:
\begin{align*}
  \mathit{Cost}_{\mathrm{total}} &= \textstyle\sum \mathit{Cost}_{\mathrm{superstep}}, &
  \mathit{Cost}_{\mathrm{superstep}} &= \mathit{Comp}_{\mathrm{total}} + \mathit{Comm}_{\mathrm{total}}.
\end{align*}
Computation cost is the sum of local computation times in the children (parallel, hence combined with $\max$) and of the local computation in the master. Addition of both terms realises the hypothesis that the master's local work may not overlap with the work of its children. Communication cost is split into the cost of performing the scatter and the gather operations:
\begin{align*}
  \mathit{Cost}_{\mathrm{superstep}} &= \mathit{Comm}_{\mathrm{scat}} + \max \mathit{Comp}_{\mathrm{children}} + \mathit{Comm}_{\mathrm{gath}} + \mathit{Comp}_{\mathrm{master}},\\
  \mathit{Comm}_{\mathrm{scat}} &= \mathit{Words}_{\mathrm{scat}} \times \mathit{Bwidth}_{\mathrm{scat}} + \mathit{Sync},\\
  \mathit{Comm}_{\mathrm{gath}} &= \mathit{Words}_{\mathrm{gath}} \times \mathit{Bwidth}_{\mathrm{gath}} + \mathit{Sync},\\
  \mathit{Comp}_{\mathrm{node}} &= \mathit{Instructions} / \mathit{ClkRate}.
\end{align*}
Communication costs are thus estimated by a linear term similar to BSP's $g\times h + L$, based on machine parameters for (the inverse of) bandwidth and for synchronisation between the master and its children. As always, local computation cost is the number of executed instructions divided by the processing speed. The machine-dependent parameters can be made as abstract or concrete as desired. In other words, theoretical SGL algorithms can be investigated with respect to their asymptotic complexity classes, while portable concrete SGL algorithms can be analysed for more precise cost formulae using bytecode-like instruction counts and normalised communication parameters. Finally, SGL programs can be compiled and measured for actual performance on a given architecture.

We remind the reader that, like the machine architecture, all the cost formulae above are intended to be recursive: the local computation cost on a child node can itself be an SGL algorithm cost if the machine structure is such.

\subsection{Parameters and detailed cost formulae}

In contrast to Multi-BSP \citep{Valiant2011}, concrete cost estimations and measurements use the following machine and algorithm parameters. Each level of a hierarchy can have different parameter values.
\begin{itemize}
  \item $l$: the latency to perform a scatter or gather synchronisation, i.e., the time to execute a 1-bit scatter or a 1-bit gather.
  \item $g$: the gap; $\gdown$ is the time interval for transmitting one word from the master to its children, and $\gup$ from the children to their master. We use a single $g$ in the case of symmetric communication costs.
  \item $k$: the number of words to transmit; $\kdown$ is the number of words that the master scatters to its children, and $\kup$ the number of words that the master gathers from its children.
  \item $w$: the work, or number of local operations performed by processors; $w_0$ is the master's work and $w_i$ ($i=1..p$) the work of child $i$. A parameter $w$ without index refers to a local quantity.
  \item $c$: the computation speed of processors; $c_0$ is the time interval for performing a unit of work on the master and $c_i$ ($i=1..p$) the time interval for performing a unit of work on child processor $i$. A parameter $c$ without index refers to a local quantity.
  \item $p$: the number of children of a master.
\end{itemize}
In general the cost formulae are
\begin{align*}
  \mathit{Cost}_{\mathrm{Master}} &= \max_{i=1..p}\bigl(\mathit{Cost}_{\mathrm{child}_i}\bigr) + w_0\times c_0 + \kdown\times\gdown + \kup\times\gup + 2l,\\
  \mathit{Cost}_{\mathrm{Worker}} &= w_i\times c_i,
\end{align*}
which clearly covers the possibility of a heterogeneous architecture. Typically, but not necessarily, communication costs are symmetric:
\[
  \mathit{Cost}_{\mathrm{superstep}} = w\times c + \Bigl[\max_{i=1..p}\bigl(\mathit{Cost}_{\mathrm{child}_i}\bigr) + (\kdown+\kup)\times g + 2l\Bigr].
\]

\section{Programming model}\label{sec:prog}

SGL's precise definition is an imperative language, for which we now give an operational semantics.

\subsection{Operational semantics}

We enrich Winskel's basic imperative language IMP \citep{Winskel1993} to yield our deterministic parallel programming language, SGL.

\subsubsection{Syntax}

\paragraph{Values.} Values are integers, booleans, and arrays (vectors) built of them. Vectors of vectors are necessary for building blocks of work to be scattered among workers.
\begin{itemize}
  \item $n$ ranges over numbers $\sort{Nat}$.
  \item $\langle n_1, n_2, \dots, n_\ell\rangle$ ranges over arrays of numbers $\sort{Vec}$, where $\ell\in\sort{Nat}$ is the length of the array.
  \item $\langle v_1, v_2, \dots, v_\ell\rangle$ ranges over arrays of arrays $\sort{VecVec}$.
\end{itemize}
The scatter operation takes a vector of vectors in the master and distributes it to the workers/children. The gather operation inverts this process.

\paragraph{Locations.} Imperative variables are abstractions of memory positions and are called \emph{locations}. They are many-sorted, like the language's values.
\begin{itemize}
  \item $X$ ranges over scalar locations $\sort{NatLoc}$, i.e., names of memory elements that store numbers. $X_{i=\mathit{pid}}$ denotes master/children locations; $X$ without index denotes the master location.
  \item $\vv{V}$ ranges over vectorial locations $\sort{VecLoc}$, i.e., names of memory elements that store arrays. $\vv{V}_{i=\mathit{pid}}$ denotes master/children locations; $\vv{V}$ without index denotes the master location.
  \item $\ww{W}$ ranges over vector-of-vector locations $\sort{VVecLoc}$, i.e., names of memory elements that store arrays of arrays.
\end{itemize}

\paragraph{Operations.}
\begin{itemize}
  \item $\odot$ denotes the binary operations $+\mid-\mid*\mid/$;
  \item $[\,]$ denotes element access in a vector;
  \item $\transfer$ denotes data to be transferred between a master and its children (i.e., an array of arrays).
\end{itemize}

\paragraph{Expressions.} Expressions are relatively standard, with the convenience of scalar-to-vector (sequential) operations.
\begin{itemize}
  \item $a$ ranges over scalar arithmetic expressions:\quad
    $\sort{Aexp} ::= n \mid X \mid a\odot a \mid \vv{V}[a]$
  \item $b$ ranges over scalar boolean expressions:\quad
    $\sort{Bexp} ::= \kw{true} \mid \kw{false} \mid a=a \mid a\le a \mid \neg b \mid b\wedge b \mid b\vee b$
  \item $v$ ranges over vectorial expressions:\quad
    $\sort{Vexp} ::= \langle a_1, a_2, \dots, a_\ell\rangle \mid \vv{V} \mid v\odot a \mid v\odot v \mid \ww{W}[a]$
  \item $w$ ranges over vector-of-vector expressions:\quad
    $\sort{VVexp} ::= \langle v_1, v_2, \dots, v_\ell\rangle \mid \ww{W}$
\end{itemize}

\paragraph{Statements.} The language's statements include the classical sequential constructs together with SGL's three primitives: \textbf{scatter}, \textbf{pardo} and \textbf{gather}. The exact meaning of the parallel statements is defined by the semantic rules below; the definition of the other statements can be found in \citet{LiHains2010}. Commands $c$ range over
\[
\begin{array}{r@{\ }l}
  \sort{Com} ::= & \kw{skip} \mid X := a \mid \vv{V} := v \mid \ww{W} := w \mid c\,;\,c \\
  \mid & \kw{if}\ b\ \kw{then}\ c\ \kw{else}\ c \mid \kw{for}\ X\ \kw{from}\ a\ \kw{to}\ a\ \kw{do}\ c \\
  \mid & \kw{scatter}\ w\ \kw{to}\ \vv{V} \mid \kw{scatter}\ v\ \kw{to}\ X \\
  \mid & \kw{gather}\ \vv{V}\ \kw{to}\ \ww{W} \mid \kw{gather}\ X\ \kw{to}\ \vv{V} \\
  \mid & \kw{pardo}\ c \mid \kw{if}\ \kw{master}\ c\ \kw{else}\ c.
\end{array}
\]

\subsubsection{Environments}

States (or environments) are maps from imperative variables (locations) to values of the corresponding sort. Like values, they are many-sorted, and we use the following notations and definitions for them. Functions in states $\Sigma$:
\begin{itemize}
  \item $\sigma : \sort{NatLoc}\to\sort{Nat}$, thus $\sigma(X)\in\sort{Nat}$;
  \item $\sigma : \sort{VecLoc}\to\sort{Vec}$, thus $\sigma(\vv{V})\in\sort{Vec}$;
  \item $\sigma : \sort{VVecLoc}\to\sort{VecVec}$, thus $\sigma(\ww{W})\in\sort{VecVec}$.
\end{itemize}
Here $\mathit{pos}\in\sort{Pos}\subseteq\sort{Nat}$ is what we call the (relative) \emph{position}: $\mathit{pos}=0$ denotes the master position (as above), and $\mathit{pos}=i\in\{1..p\}$ denotes the position of the $i$-th child. It is the recursive analogue of BSP's (or MPI's) \texttt{pid}.
\begin{itemize}
  \item $\sigma : \sort{NatLoc}\to\sort{Pos}\to\sort{Nat}$, thus $\sigma(X_{\mathit{pos}}) = \sigma_{\mathit{pos}}(X)\in\sort{Nat}$;
  \item $\sigma : \sort{VecLoc}\to\sort{Pos}\to\sort{Vec}$, thus $\sigma(\vv{V}_{\mathit{pos}}) = \sigma_{\mathit{pos}}(\vv{V})\in\sort{Vec}$.
\end{itemize}

\paragraph{Rules for vectorial expressions.} The semantics of vector expressions is standard and deserves no special explanation.
\begin{mathpar}
  \inferrule{\forall i=1..\ell\quad \cfg{a_i}{} \to n_i}
            {\cfg{\langle a_1, a_2, \dots, a_\ell\rangle}{} \to \langle n_1, n_2, \dots, n_\ell\rangle}
  \and
  \inferrule{ }{\cfg{\vv{V}}{} \to \sigma(\vv{V})}
  \and
  \inferrule{\cfg{\ww{W}}{} \to \sigma(\ww{W}) = \langle v_1, \dots, v_\ell\rangle \\ \cfg{a}{} \to n}
            {\cfg{\ww{W}[a]}{} \to v_n}
  \\
  \inferrule{\cfg{v}{} \to \langle n'_1, \dots, n'_\ell\rangle \\ \cfg{a}{} \to n \\ \forall i=1..\ell\quad \cfg{n'_i\odot n}{} \to n_i}
            {\cfg{v\odot a}{} \to \langle n_1, \dots, n_\ell\rangle}
  \\
  \inferrule{\cfg{v_1}{} \to \langle n'_1, \dots, n'_\ell\rangle \\ \cfg{v_2}{} \to \langle n''_1, \dots, n''_\ell\rangle \\ \forall i=1..\ell\quad \cfg{n'_i\odot n''_i}{} \to n_i}
            {\cfg{v_1\odot v_2}{} \to \langle n_1, \dots, n_\ell\rangle}
\end{mathpar}
(The lengths of $v_1$ and $v_2$ must be equal.) The rules for the other expressions are written similarly.

\paragraph{Primitive parallel statements.} In the following rules, $\numChd$ is the number of children of the processor.

\smallskip\noindent\emph{Scatters.}
\begin{mathpar}
  \inferrule{\cfg{w}{} \to \langle v_1, \dots, v_\ell\rangle \\ \forall i=1..\numChd\quad \cfg{\vv{V}_i \transfer v_i}{} \to \sigma'_i}
            {\cfg{\kw{scatter}\ w\ \kw{to}\ \vv{V}}{} \to \sigma'}
  \and
  \inferrule{\cfg{v}{} \to \langle n_1, \dots, n_\ell\rangle \\ \forall i=1..\numChd\quad \cfg{X_i \transfer n_i}{} \to \sigma'_i}
            {\cfg{\kw{scatter}\ v\ \kw{to}\ X}{} \to \sigma'}
\end{mathpar}
\emph{Gathers.}
\begin{mathpar}
  \inferrule{\cfg{\ww{W} \transfer \langle \vv{V}_1, \vv{V}_2, \dots, \vv{V}_{\numChd}\rangle}{} \to \sigma'}
            {\cfg{\kw{gather}\ \vv{V}\ \kw{to}\ \ww{W}}{} \to \sigma'}
  \and
  \inferrule{\cfg{\vv{V} \transfer \langle X_1, X_2, \dots, X_{\numChd}\rangle}{} \to \sigma'}
            {\cfg{\kw{gather}\ X\ \kw{to}\ \vv{V}}{} \to \sigma'}
\end{mathpar}
\emph{Parallel.}
\begin{mathpar}
  \inferrule{\forall i=1..\numChd\quad \langle c,\ \sigma_i\rangle \to \sigma'_i}
            {\cfg{\kw{pardo}\ c}{} \to \sigma'}
  \and
  \inferrule{\cfg{\numChd=0}{} \to \kw{true} \\ \cfg{c_2}{} \to \sigma'}
            {\cfg{\kw{if}\ \kw{master}\ c_1\ \kw{else}\ c_2}{} \to \sigma'}
  \and
  \inferrule{\cfg{\numChd=0}{} \to \kw{false} \\ \cfg{c_1}{} \to \sigma'}
            {\cfg{\kw{if}\ \kw{master}\ c_1\ \kw{else}\ c_2}{} \to \sigma'}
\end{mathpar}

\section{Experimentation}\label{sec:exp}

Now that the language's cost model and operational semantics have been defined, we summarise some performance measurements and analyse them with respect to the cost model.

The local processing speed $c$ and the network parameters $l$, $g$ can be measured on a chosen machine once the hierarchy of processors and the benchmarking setup have been defined. For a given algorithm, we can also analyse theoretically, estimate or count the $w$ quantities.

We have performed experiments to observe how the measured computation time of an SGL algorithm compares with the cost model's prediction. The overall system has the following description:

\begin{center}\small
\begin{tabular}{@{}ll@{}}
  \textbf{Nodes}   & 16 computing nodes (2 CPUs/node)\\
  \textbf{CPU}     & Intel Xeon E5440 (quad-core, 2.83\,GHz)\\
  \textbf{RAM}     & 512\,GB in total (4\,GB/core, DDR2 667)\\
  \textbf{Network} & 4X DDR InfiniBand switches (16\,Gbit/s)
\end{tabular}
\end{center}

\begin{figure}[htbp]
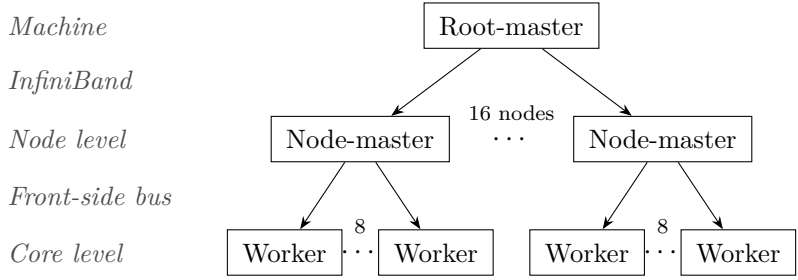

  \centering\figClusterMachine
  \caption{The two-level SGL machine for EXQIM's cluster.}\label{fig:cluster}
\end{figure}

\noindent We build a two-level SGL abstract machine to represent this physical machine (\cref{tab:machine,fig:cluster}).

\begin{table}[htbp]
  \centering\small
  \caption{Two-level SGL abstract machine for the experimental cluster.}\label{tab:machine}
  \begin{tabular}{@{}lll@{}}
    \toprule
    Unit & Children & Communication \\
    \midrule
    Root-master & 16 nodes & InfiniBand \\
    Node-master & 8 cores  & Front-side bus \\
    Worker      & 0        & N/A \\
    \bottomrule
  \end{tabular}
\end{table}

\subsection{Parameter measurement}

The CPUs of the computing nodes are clocked at 2.83\,GHz, so that $c = 0.353$\,ns/op for each of them.

At node level, we use the collective functions \texttt{MPI\_Barrier} for $L$, \texttt{MPI\_Scatterv} for $\gdown$ and \texttt{MPI\_Gatherv} for $\gup$ of SGI's Message Passing Toolkit (MPT 2.04). In \cref{tab:node}, the first four lines were used in our experiments of SGL on a hierarchical architecture, and the last four lines are used for comparison with BSP on a flat architecture.

\begin{table}[htbp]
  \centering\small
  \caption{Node-level parameters. With SGI's MPI implementation the cost of $g$ increases with the number of processes, and \texttt{MPI\_Gatherv} has a minimum around 2\,ns/32\,b.}\label{tab:node}
  \begin{tabular}{@{}rcrrr@{}}
    \toprule
    Number of procs & Nodes $\times$ cores & $L$ ($\mu$s) & $\gdown$ (ns/32\,b) & $\gup$ (ns/32\,b) \\
    \midrule
      2 & $2\times1$  & 1.48 & 1.38 & 2.15 \\
      4 & $4\times1$  & 2.85 & 1.69 & 2.00 \\
      8 & $8\times1$  & 4.37 & 1.89 & 2.05 \\
     16 & $16\times1$ & 5.96 & 2.04 & 2.09 \\
    \midrule
     32 & $16\times2$ & 7.62 & 2.14 & 2.09 \\
     64 & $16\times4$ & 7.93 & 2.63 & 2.11 \\
     96 & $16\times6$ & 8.81 & 2.88 & 2.13 \\
    128 & $16\times8$ & 9.89 & 3.01 & 2.77 \\
    \bottomrule
  \end{tabular}
\end{table}

\begin{figure}[htbp]
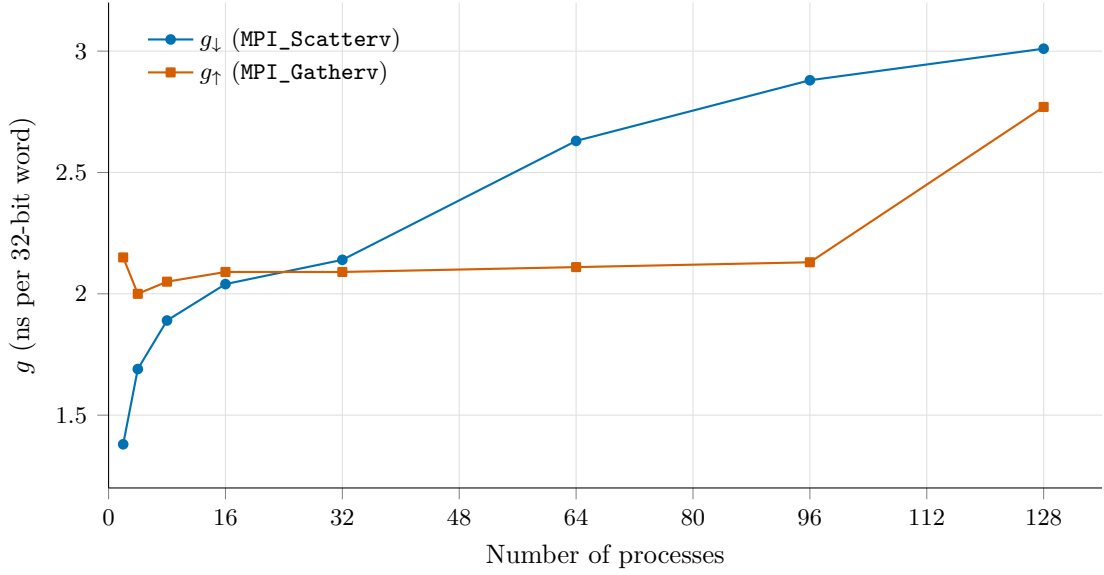

  \centering\figGapMPI
  \caption{Measurement of $g$ in MPI (data of \cref{tab:node}).}\label{fig:gap}
\end{figure}

At the core (worker) level, we use OpenMP's barrier for $L$ and the C function \texttt{memcpy} for $g$. Instead of directly transferring pointers to data, we use \texttt{memcpy()} to copy data into different memory regions, to avoid concurrent access between CPU cores. \Cref{tab:core} shows the measured core-level parameters.

\begin{table}[htbp]
  \centering\small
  \caption{Core-level parameters.}\label{tab:core}
  \begin{tabular}{@{}lrr@{}}
    \toprule
    Machine & $L$ ($\mu$s) & $g$ (ns/32\,b) \\
    \midrule
    2 cores & 12.08 & 0.59 \\
    4 cores & 25.64 & 0.59 \\
    6 cores & 37.80 & 0.59 \\
    8 cores & 52.00 & 0.59 \\
    \bottomrule
  \end{tabular}
\end{table}

According to \cref{tab:node,tab:core} (see also \cref{fig:gap}), if we used flat BSP instead of the SGL model to represent our machine, the communication cost between root-master and workers would increase by nearly 0.4\,ns/32\,b. To represent our physical machine in BSP, we would build a 128-processor abstract machine using MPI. Under this architecture, the value of $g$ would be $\max(3.01, 2.77) = 3.01$\,ns/32\,b. In SGL, we instead built a 16-processor network at node level using MPI and an 8-processor network at core level using OpenMP. Thus the value of $\gdown$ is $2.04 + 0.59 = 2.63$\,ns/32\,b and the value of $\gup$ is $2.09 + 0.59 = 2.68$\,ns/32\,b. This is a medium-scale but concrete example of the practical gains obtained from a hierarchical model that better fits the physical architecture.

\subsection{Algorithms}

We have tested SGL variants of some of the most important basic parallel algorithms: parallel reduction (with the product operation), parallel prefix reduction, also called scan (with the sum operation), and a sorting algorithm. For each one we wrote an SGL algorithm, implemented it by hand using the MPI/OpenMP operations mentioned above, and then compared the model's predicted and observed run times for increasing data sizes. Finally, we computed speed-up and parallel-efficiency values.

The three tests illustrate SGL's relative programming simplicity on a relevant set of algorithms, give convincing evidence of the cost model's quality, and provide initial evidence that SGL programming has no intrinsic cost overhead.

\subsubsection{Parallel reduction}

\begin{figure}[htbp]
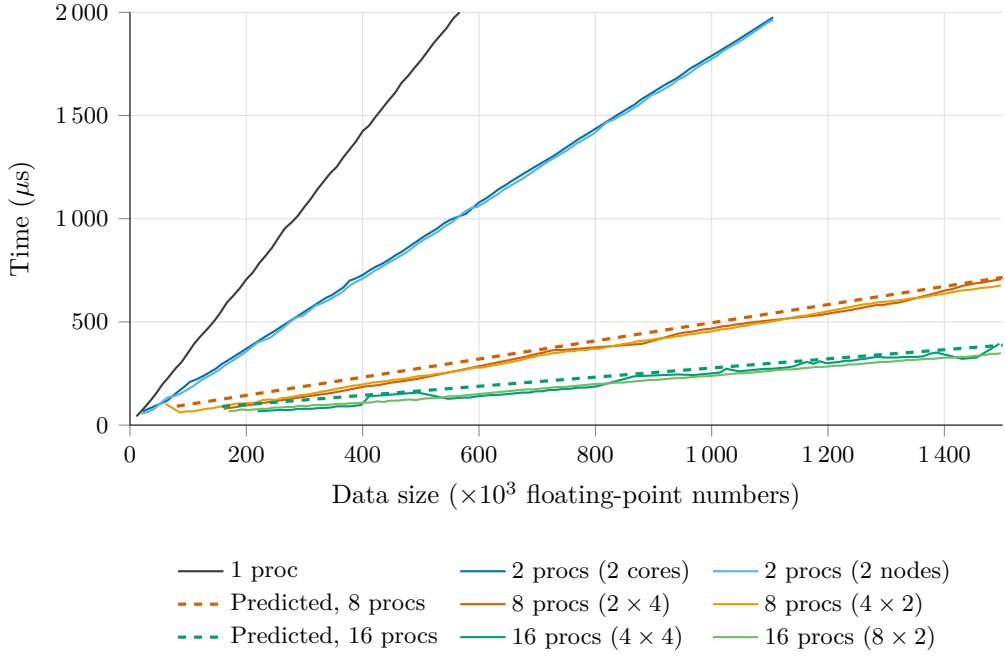

  \centering\perfplot{reduction}{1500}{2000}
  \caption{Predicted and actual execution times for SGL reduction.}\label{fig:reduction}
\end{figure}

In \cref{alg:reduction}, each worker computes the product of its local scalar numbers. After that, the master fetches the computed products from its children and calculates the final product (annotations on the right are cost estimates for the corresponding lines).

\begin{algorithm}[htbp]
\caption{Parallel reduction}\label{alg:reduction}
\begin{algorithmic}[1]
  \Proc{Reduction}{IN $\vv{\mathit{src}}$, OUT $\mathit{res}$}
  \IfMaster
    \ParDo
      \State Reduction($\vv{\mathit{src}}$, $\mathit{res}$); \Cost{\mathit{Reduction}_{\mathrm{child}_i}}
    \EndParDo
    \State \Gather{$\mathit{res}$}{$\vv{\mathit{lst}}$}; \Cost{p\times\gup + l}
    \State Product($\vv{\mathit{lst}}$, $\mathit{res}$); \Cost{O(p)}
  \Else
    \State Product($\vv{\mathit{src}}$, $\mathit{res}$); \Cost{O(n)}
  \EndIf
  \ProcEnd
  \Statex
  \Proc{Product}{IN $\vv{\mathit{src}}$, OUT $\mathit{res}$}
  \State $\mathit{res} := 1$;
  \For{$i$ \textbf{from} 1 \textbf{to} $\mathrm{len}(\vv{\mathit{src}})$}
    \State $\mathit{res} := \mathit{res} * \vv{\mathit{src}}[i]$;
  \EndFor
  \ProcEnd
\end{algorithmic}
\end{algorithm}

Line~3 is a recursive call to the algorithm, and line~8, the no-children case, is a local sequential loop. Product operates on $\vv{\mathit{src}}$, whose elements are the $\mathit{res}$ outputs of the children. The cost of the superstep is (subscripts $\langle\cdot\rangle$ refer to the pseudo-code lines):
\begin{align*}
  \mathit{Cost}_{\mathrm{Master}} &= \max_{i=1..p}\bigl(\mathit{Reduction}_{\mathrm{child}_i}\langle3\rangle\bigr) + O(p)\langle6\rangle\times c + p\langle5\rangle\times\gup + l,\\
  \mathit{Cost}_{\mathrm{Worker}} &= O(n_{\mathrm{worker}})\langle8\rangle\times c.
\end{align*}
The measurements (\cref{fig:reduction}) show a nearly perfect match between measured performance and predicted performance, the latter being obtained from the cost formula with machine parameters measured independently of this algorithm.

\subsubsection{Parallel scan}

\begin{figure}[htbp]
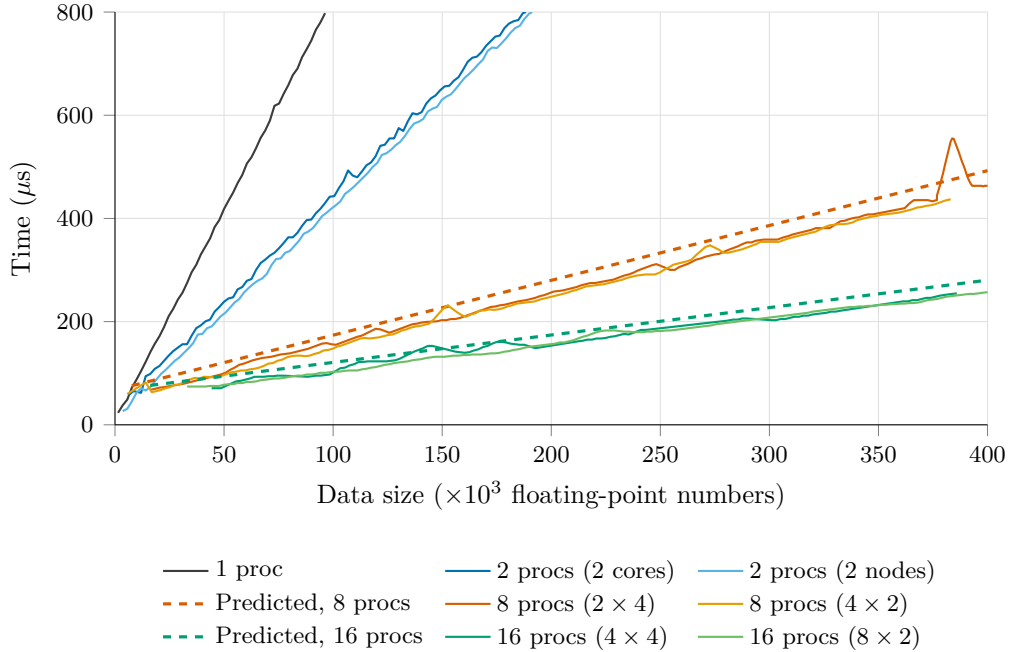

  \centering\perfplot{scan}{400}{800}
  \caption{Predicted and actual execution times for SGL scan.}\label{fig:scan}
\end{figure}

We perform the parallel scan algorithm (\cref{alg:scan}) in two steps:
\begin{enumerate}
  \item Each worker performs a sequential scan. Then the master fetches the last computed element from each child. After that, it performs a sequential scan of the fetched data.
  \item The master scatters the results among its children. Then each child adds the received value to its computed data. Finally, the children hold the final result.
\end{enumerate}
The pseudo-code is given in \cref{alg:scan}, and the cost of the superstep is
\begin{align*}
  \mathit{Cost}_{\mathrm{Master}} ={}& \max_{i=1..p}\bigl(\mathit{Step1}_{\mathrm{child}_i} + O(1)\times c_i\bigr) + \max_{i=1..p}\bigl(\mathit{Step2}_{\mathrm{child}_i} + O(n_{\mathrm{child}_i})\times c_i\bigr)\\
  &+ \bigl(O(p) + O(p-1)\bigr)\times c + p\times\gup + p\times\gdown + 2\times l,\\
  \mathit{Cost}_{\mathrm{Worker}} ={}& O(n_{\mathrm{worker}})\times c.
\end{align*}
Again, the pseudo-code and the measurements (\cref{fig:scan}) support our claims that SGL is simple to use and that its performance model is reliable. Several observations can be made from this graph: the performance prediction is correct to within a few percent, and the discrepancy between prediction and observation increases with communication cost or with the number of processing units. Moreover, configurations with more cores per processor exhibit larger variability due to the local shared memory.

\begin{algorithm}[htbp]
\caption{Parallel scan}\label{alg:scan}
\begin{algorithmic}[1]
  \Proc{Step1}{IN $\vv{\mathit{src}}$, OUT $\vv{\mathit{mid}}$}
  \IfMaster
    \ParDo
      \State Step1($\vv{\mathit{src}}$, $\vv{\mathit{mid}}$); \Cost{\mathit{Step1}_{\mathrm{child}_i}}
      \State $x := \vv{\mathit{mid}}[\mathrm{len}(\vv{\mathit{mid}})]$; \Cost{O(1)_{\mathrm{child}_i}}
    \EndParDo
    \State \Gather{$x$}{$\vv{\mathit{mid}}$}; \Cost{p\times\gup + l}
    \State ShiftRight($\vv{\mathit{mid}}$); \Cost{O(p)}
    \State LocalScan($\vv{\mathit{mid}}$, $\vv{\mathit{mid}}$); \Cost{O(p)}
  \Else
    \State LocalScan($\vv{\mathit{src}}$, $\vv{\mathit{mid}}$); \Cost{O(n)}
  \EndIf
  \ProcEnd
  \Statex
  \Proc{Step2}{IN $\vv{\mathit{mid}}$, OUT $\vv{\mathit{res}}$}
  \IfMaster
    \State \Scatter{$\vv{\mathit{mid}}$}{$x$}; \Cost{p\times\gdown + l}
    \ParDo
      \State $\vv{\mathit{res}} := \vv{\mathit{mid}} + x$; \Cost{O(n_{\mathrm{child}_i})}
      \State Step2($\vv{\mathit{res}}$, $\vv{\mathit{res}}$); \Cost{\mathit{Step2}_{\mathrm{child}_i}}
    \EndParDo
  \Else
    \State \textbf{skip};
  \EndIf
  \ProcEnd
  \Statex
  \Proc{LocalScan}{IN $\vv{\mathit{src}}$, OUT $\vv{\mathit{res}}$}
  \State $\vv{\mathit{res}}[1] := \vv{\mathit{src}}[1]$;
  \For{$i$ \textbf{from} 2 \textbf{to} $\mathrm{len}(\vv{\mathit{src}})$}
    \State $\vv{\mathit{res}}[i] := \vv{\mathit{res}}[i-1] + \vv{\mathit{src}}[i]$;
  \EndFor
  \ProcEnd
  \Statex
  \Proc{ShiftRight}{INOUT $\vv{\mathit{lst}}$}
  \For{$i$ \textbf{from} $\mathrm{len}(\vv{\mathit{lst}})$ \textbf{downto} 2}
    \State $\vv{\mathit{lst}}[i] := \vv{\mathit{lst}}[i-1]$;
  \EndFor
  \State $\vv{\mathit{lst}}[1] := 0$;
  \ProcEnd
\end{algorithmic}
\end{algorithm}

We now experiment with a BSP sorting algorithm in SGL.

\subsubsection{Parallel sorting}

Our parallel sorting algorithm (\cref{alg:psrs1,alg:psrs2}) is based on parallel sorting by regular sampling, PSRS \citep{Shi1992}, a partition-based algorithm. We could also use the algorithm of \citet{Helman1998} to improve performance. We implement this algorithm in five steps:
\begin{enumerate}
  \item Each worker performs a local sort and selects $p$ (the number of workers) samples, which are gathered onto the root-master.
  \item The root-master performs a local sort of the $p^2$ gathered samples. Then it picks $p-1$ almost equally spaced pivots from the sorted samples.
  \item The root-master broadcasts these pivots to all workers. After that, each worker produces $p$ partitions of its local data using the $p-1$ pivots. Partition $i$ holds the values that should now be moved to worker $i$.
  \item The master gathers the partitions that are not already in place.
  \item The master scatters the gathered partitions to its children according to the partitions' indices. Then each worker performs a local merge of the received partitions.
\end{enumerate}

\begin{algorithm}[htbp]
\caption{Parallel sorting by regular sampling, steps 0--3}\label{alg:psrs1}
\begin{algorithmic}[1]
  \Statex \textbf{Step0}
  \State initialise $\mathit{pid} := 1$ in each worker;
  \State use parallel scan (\cref{alg:scan}) to compute the $\mathit{pid}$ of each worker;
  \State set $\mathit{lowerPid}$, $\mathit{upperPid}$ and $\mathit{maxPid}$ in masters;
  \Statex
  \Proc{Step1}{IN $\vv{\mathit{arr}}$, OUT $\vv{\mathit{sam}}$}
  \IfMaster
    \ParDo
      \State Step1($\vv{\mathit{arr}}$, $\vv{\mathit{sam}}$);
    \EndParDo
    \State \Gather{$\vv{\mathit{sam}}$}{$\ww{\mathit{tmp}}$};
    \State $\vv{\mathit{sam}} :=$ Concatenate($\ww{\mathit{tmp}}$);
  \Else
    \State QuickSort($\vv{\mathit{arr}}$);
    \State SelectSamples($\vv{\mathit{arr}}$, $\vv{\mathit{sam}}$);
  \EndIf
  \ProcEnd
  \Statex
  \Proc{Step2}{IN $\vv{\mathit{sam}}$, OUT $\vv{\mathit{pvt}}$}
  \State QuickSort($\vv{\mathit{sam}}$);
  \State PickPivots($\vv{\mathit{sam}}$, $\vv{\mathit{pvt}}$);
  \ProcEnd
  \Statex
  \Proc{Step3}{IN $\vv{\mathit{arr}}$, IN $\vv{\mathit{pvt}}$, OUT $\ww{\mathit{blk}}$}
  \IfMaster
    \For{$i$ \textbf{from} 1 \textbf{to} $\numChd$}
      \State $\ww{\mathit{tmp}}[i] := \vv{\mathit{pvt}}$;
    \EndFor
    \State \Scatter{$\ww{\mathit{tmp}}$}{$\vv{\mathit{pvt}}$};
    \ParDo
      \State Step3($\vv{\mathit{arr}}$, $\vv{\mathit{pvt}}$, $\ww{\mathit{blk}}$);
    \EndParDo
  \Else
    \State BuildPartitions($\vv{\mathit{arr}}$, $\vv{\mathit{pvt}}$, $\ww{\mathit{blk}}$);
  \EndIf
  \ProcEnd
\end{algorithmic}
\end{algorithm}

\begin{algorithm}[htbp]
\caption{Parallel sorting by regular sampling, steps 4--5}\label{alg:psrs2}
\begin{algorithmic}[1]
  \Proc{Step4}{IN $\ww{\mathit{blk}}$, OUT $\ww{\mathit{stay}}$, OUT $\ww{\mathit{move}}$}
  \IfMaster
    \ParDo
      \State Step4($\ww{\mathit{blk}}$, $\ww{\mathit{stay}}$, $\ww{\mathit{move}}$);
    \EndParDo
    \For{$i$ \textbf{from} 1 \textbf{to} $\mathit{maxPid}$}
      \State \Gather{$\ww{\mathit{move}}[i]$}{$\ww{\mathit{tmp}}$};
      \If{$i < \mathit{lowerPid}$ \textbf{or} $i > \mathit{upperPid}$}
        \State $\ww{\mathit{move}}[i] :=$ Concatenate($\ww{\mathit{tmp}}$);
      \Else
        \State $\ww{\mathit{stay}}[i] :=$ Concatenate($\ww{\mathit{tmp}}$);
      \EndIf
    \EndFor
  \Else
    \For{$i$ \textbf{from} 1 \textbf{to} $\mathit{maxPid}$}
      \If{$i = \mathit{pid}$}
        \State $\ww{\mathit{stay}}[\mathit{pid}] := \ww{\mathit{blk}}[\mathit{pid}]$;
      \Else
        \State $\ww{\mathit{move}}[i] := \ww{\mathit{blk}}[i]$;
      \EndIf
    \EndFor
  \EndIf
  \ProcEnd
  \Statex
  \Proc{Step5}{IN $\ww{\mathit{stay}}$, IN $\ww{\mathit{move}}$, OUT $\vv{\mathit{arr}}$}
  \IfMaster
    \State \Scatter{Bundle($\ww{\mathit{move}}$)}{$\vv{\mathit{arr}}$};
    \ParDo
      \State $\ww{\mathit{tmp}} :=$ Unbundle($\vv{\mathit{arr}}$);
      \For{$i$ \textbf{from} $\mathit{lowerPid}$ \textbf{to} $\mathit{upperPid}$}
        \State $\ww{\mathit{move}}[i] :=$ Concatenate($\ww{\mathit{tmp}}[i]$, $\ww{\mathit{stay}}[i]$);
      \EndFor
      \State Step5($\ww{\mathit{stay}}$, $\ww{\mathit{move}}$, $\vv{\mathit{arr}}$);
    \EndParDo
  \Else
    \State $\vv{\mathit{arr}} :=$ MergeSort($\ww{\mathit{move}}[\mathit{pid}]$);
  \EndIf
  \ProcEnd
\end{algorithmic}
\end{algorithm}

\citet{Sujithan1996} showed that the computational cost of this algorithm is $2\frac{n}{p}\bigl(\log n - \log p + \frac{p^3}{n}\log p\bigr)$ and its communication cost is $g\frac{1}{p}\bigl(p^2(p-1)+n\bigr) + 4l$ in BSP. Thus the total cost of this algorithm implemented in SGL is
\[
  2\frac{n}{p}\Bigl(\log n - \log p + \frac{p^3}{n}\log p\Bigr)\times c + \frac{1}{p}\bigl(p^2(p-1)+n\bigr)\times G + 4\times L,
\]
where $G$ is the sum of the $g$ of all levels and $L$ is the sum of the $l$ of all levels.

Our performance prediction is based on $\mathrm{SeqSort}(n) = O(n\log n)$ for the sequential sorting algorithm and $O(n/p)$ for the data to be transferred from one node to another after the local sort. Our implementation includes a node-to-node horizontal communication optimisation, which we have begun to investigate formally and will add to an SGL compiler.

\begin{figure}[htbp]
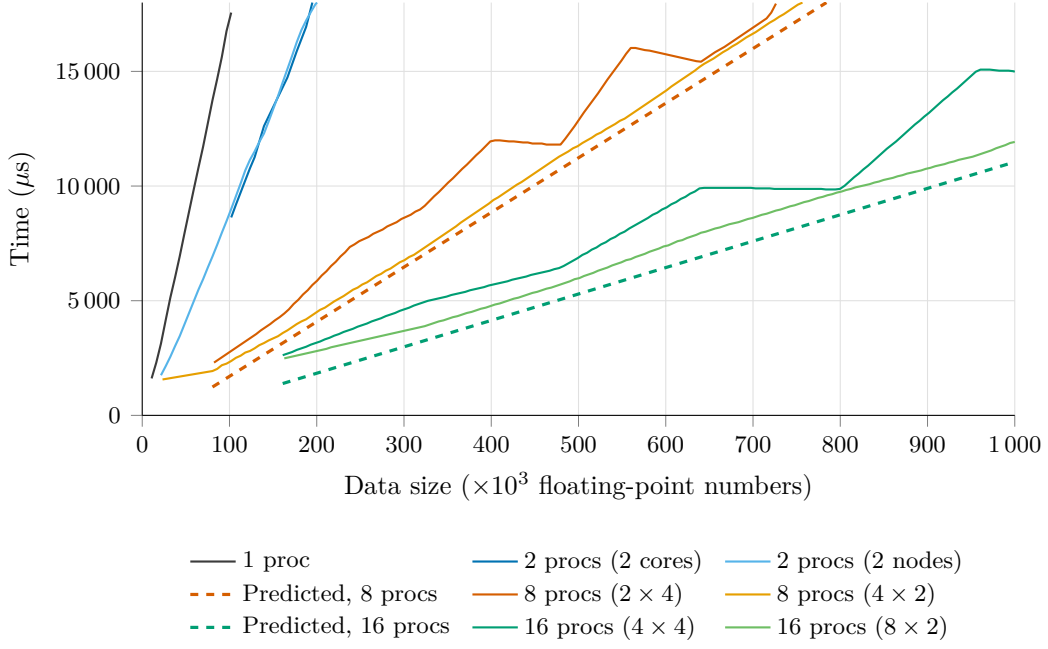

  \centering\perfplot{psrs}{1000}{18000}
  \caption{Predicted and actual execution times for SGL parallel sorting by regular sampling.}\label{fig:psrs}
\end{figure}

The measurements shown in \cref{fig:psrs} lead to the same observations and explanations as above for the parallel scan. Higher local computation costs lead to a larger prediction error; in particular, our theoretical estimate is lower than the observations.

The pseudo-code shows that our primitive algorithms are easily implemented in SGL. Future work will investigate SGL programming for more general algorithms. We now provide some quantification of SGL's capacity to generate large speed-ups.

\subsection{Speed-up and efficiency}

We tested speed-up and efficiency using the reduction algorithm of \cref{alg:reduction}. The formulae we used are
\[
  \mathit{Speedup}_p = \frac{\mathit{ExecutionTime}_1}{\mathit{ExecutionTime}_p},
  \qquad
  \mathit{Efficiency}_p = \frac{\mathit{Speedup}_p}{p},
\]
where $p$ is the number of nodes (resp.\ cores) used.

First, we fixed the size of the input data at 2\,560\,000 floating-point numbers and the number of cores at four per node, and varied the number of nodes from 1 to 16. The results are given in \cref{tab:nodescale}. We then fixed the size of the input data at 9\,600\,000 floating-point numbers and the number of nodes at 16, and varied the number of cores per node from 1 to 4. The results are given in \cref{tab:corescale}.

\begin{table}[htbp]
  \centering\small
  \caption{Node-level scale-out (four cores per node, 2\,560\,000 floats).}\label{tab:nodescale}
  \begin{tabular}{@{}lrrrrr@{}}
    \toprule
    Number of procs & 4 & 8 & 16 & 32 & 64 \\
    Nodes           & 1 & 2 & 4  & 8  & 16 \\
    \midrule
    Speed-up        & 1 & 1.999 & 3.974 & 7.560 & 13.815 \\
    Efficiency      & 1 & 0.999 & 0.994 & 0.945 & 0.863 \\
    \bottomrule
  \end{tabular}
\end{table}

\begin{table}[htbp]
  \centering\small
  \caption{Core-level scale-out (16 nodes, 9\,600\,000 floats).}\label{tab:corescale}
  \begin{tabular}{@{}lrrrr@{}}
    \toprule
    Number of procs & 16 & 32 & 48 & 64 \\
    Cores per node  & 1  & 2  & 3  & 4 \\
    \midrule
    Speed-up        & 1 & 1.965 & 2.884 & 3.852 \\
    Efficiency      & 1 & 0.983 & 0.961 & 0.963 \\
    \bottomrule
  \end{tabular}
\end{table}

\begin{figure}[htbp]
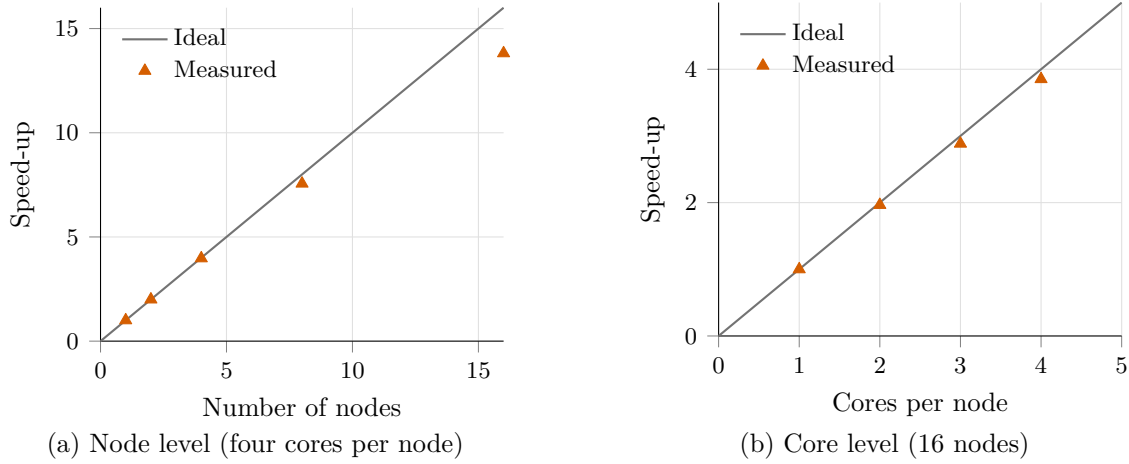

  \centering
  \begin{minipage}[t]{0.48\linewidth}\centering
    \speedupplot{(1,1)(2,1.999)(4,3.974)(8,7.560)(16,13.815)}{16}{Number of nodes}\\[-2pt]
    \small (a) Node level (four cores per node)
  \end{minipage}\hfill
  \begin{minipage}[t]{0.48\linewidth}\centering
    \speedupplot{(1,1)(2,1.965)(3,2.884)(4,3.852)}{5}{Cores per node}\\[-2pt]
    \small (b) Core level (16 nodes)
  \end{minipage}
  \caption{Measured speed-up of SGL reduction: (a) node-level scale-out (\cref{tab:nodescale}); (b) core-level scale-out (\cref{tab:corescale}).}\label{fig:speedup}
\end{figure}

Our measurements show that core-level scale-out is more efficient than node-level scale-out. The theoretical explanation is that node-level communication is more expensive than core-level communication. \Cref{fig:speedup} illustrates the two tables from the point of view of acceleration.

\section{Conclusions}\label{sec:concl}

Scatter-gather parallel operations are old, well-understood concepts, and we do not pretend that SGL's use of them is novel. What this paper has attempted to motivate and support is that BSP's advantages for parallel software can be enhanced by the recursive, hierarchical and heterogeneous machine structure of SGL, while simplifying the programming interface even further by replacing point-to-point messages with logically centralised communications. From the perspective of programming research, SGL is a proposal to:
\begin{enumerate}
  \item adapt BSML to a hierarchical machine structure;
  \item remove its general communication primitive \texttt{put} (\texttt{put} is no longer a primitive but remains a possible optimised implementation);
  \item replace \texttt{mkpar} with the scatter operation; and
  \item replace \texttt{proj} with the gather operation.
\end{enumerate}
Our initial experiments with language definition, programming and performance measurement show that SGL combines a clean semantics, simplified programming of BSP-like algorithms and dependable performance measurement. One remaining open problem is the implicit treatment of the horizontal communication needed for operations like sample sort or bucket sort \citep{Shi1992}. We are confident that the large body of knowledge on algorithms for cost-explicit models will yield many solutions to this problem. It is our opinion that SGL's lesser expressive power with respect to message-passing structures is not only necessary for software reliability but also coherent with one of the most basic concepts in computing, namely recursive structures.

\section{Future work}\label{sec:future}

Ongoing work around SGL aims at developing:
\begin{enumerate}
  \item an MPI/OpenMP implementation of a library for SGL's parallel operations;
  \item a compiler for the simple imperative SGL language described in this paper;
  \item performance prediction for this compiler, based on our performance model;
  \item optimisations based on the performance-prediction model and on the fundamental equation of modelling \citep{Barker2009}, $T_{\mathrm{total}} = T_{\mathrm{comp}} + T_{\mathrm{comm}} - T_{\mathrm{overlap}}$;
  \item a compiler from a domain-specific source application language (with no explicit parallel construct) to SGL;
  \item horizontal child-to-child communications, as implicit semantics or as optimisations of SGL program executions.
\end{enumerate}
All of the above aspects of SGL's development assume that it can be compiled and executed with a variable SGL-machine view, to more or less match the physical architecture. For example, a network of bi-processors built from quad-core processors can have one, two or three levels when viewed as an SGL computer.

At a more technical level, future work will investigate:
\begin{enumerate}
  \item horizontal and vertical load balancing for heterogeneous machines;
  \item pipelining or overlap behaviour, in the operational semantics or as optimisations;
  \item SGL support for algorithms that use such horizontal communications;
  \item including memory size in the model;
  \item extended SGL implementations with accelerators (CUDA, Cell, SSE, AVX, etc.);
  \item extended SGL implementations supporting fault tolerance.
\end{enumerate}

\section*{Acknowledgements}

The first author thanks EXQIM S.A.S.\ for an Industrial Doctoral Scholarship under the French Government's CIFRE scheme. The second author thanks W.\,F.~McColl for introducing him to the BSP model in the early 1990s. SGL is inspired by the work of colleagues who are too numerous to cite here, but special thanks are due to Fr\'ed\'eric Loulergue and colleagues for their work on BSML. Both authors thank Fr\'ed\'eric Gava for useful comments on this paper, and EXQIM for access to the SGI hardware on which we conducted our experiments.

\bibliographystyle{plainnat}
\bibliography{sgl}

\end{document}